\documentstyle[amssymb,a4,12pt]{article}

{\catcode `\@=11 \global\let\AddToReset=\@addtoreset}
\AddToReset{equation}{section}

\newtheorem{Theorem} {Theorem} [section]  \newtheorem{Lemma} [Theorem] {Lemma}

\def\scri{\hbox{${\cal J}$\kern -.645em {\raise
      .57ex\hbox{$\scriptscriptstyle (\ $}}}}
 \newcommand{\eq}[1]{(\ref{#1})}
 \newcommand{\be}{\begin{equation}}
  \newcommand{\ee}{\end{equation}} 

  \def\Reals{{\Bbb R}}

 \newcommand{\R}{\Reals}

\newcommand{\commentout}[1]{}

\newcounter{mnotecount}[section]

\newcommand{\mnote}[1]{}

\begin{document}

\title{A remark on differentiability of Cauchy horizons}

\author{Piotr T.\ Chru\'sciel\thanks{Alexander von Humboldt
    Fellow. Supported in part by the French Ministry of Foreign
    Affairs, and by the grant KBN 2 P03B 073 15.  E--mail:
    Chrusciel@Univ-Tours.Fr}
  \\ D\'epartement de Math\'ematiques \\ Facult\'e des Sciences et
  Techniques \\ Universit\'e de Tours \\ Parc de Grandmont, F-37200
  Tours, France}

\maketitle

\begin{abstract}
  In a recent paper Kr\'olak and Beem \cite{BK2} have shown
  differentiability of Cauchy horizons at all points of multiplicity
  one. In this note we give a simpler proof of this result.
\end{abstract}

\section{Introduction} 

A question of current interest is that of differentiability of various
horizons that occur in general relativity. Recall that in
\cite{ChGalloway} it was shown that there exist Cauchy horizons, as
well as black hole event horizons, which are non differentiable on a
dense set. In that reference it was also shown that
\begin{enumerate}
\item Cauchy horizons are differentiable at all interior points of
  their generators;
\item Cauchy horizons are not differentiable at all end points of
  generators of multiplicity larger than one.
\end{enumerate}
(Recall that the multiplicity of an end point of a generator is
defined as the number (perhaps infinite) of generators which end at
this point.) These results leave open the question of
differentiability of a Cauchy horizon at end points of multiplicity
one.  In a recent paper Kr\'olak and Beem \cite{BK2} have settled this
issue, showing differentiability of Cauchy horizons at those points.
In this note we give a simpler proof of this result. Actually,
motivated by the question of differentiability of black hole horizons,
we will prove differentiability of a somewhat larger class of
hypersurfaces, {\em cf.\/} Theorem \ref{T2} below.

{\bf Acknowledgments} Useful discussions with G. Galloway concerning a
previous version of this paper are acknowledged.  The author is
grateful to the Department of Mathematics of the University of
Canberra for hospitality and financial support during work on this
paper.

\def\spar{\par\smallskip\noindent} \def\mpar{\par\medskip\noindent}
\def\bpar{\par\bigskip\noindent}
\section{Statements and proofs}
Before proving our main result, Theorem \ref{T2}, we need the
following preliminary result:
\begin{Lemma}\label{L1}
  Let ${\cal U} \subset {\Bbb R}^n$ be an open set, suppose that $f
  \in C^{0, 1} (\cal U)$ and consider
  $${\cal H} = \left\{t = f(\vec x), \vec x \in \cal U \right\}.$$
  Then $\cal H$ is differentiable at $\vec x_0 \in \cal U$ if and only
  if there exists a hypersurface $T \subset {\Bbb R} \times {\Bbb
    R}^n$ such that for every sequence $\left(f(\vec x_0) + \epsilon_i
    w_i, x_0 + \epsilon_i \vec v_i\right) \in {\cal H}$ \ for which
  $\epsilon_i \rightarrow 0, w_i \rightarrow w$ and $\vec v_i
  \rightarrow \vec v$ we have $(w, \vec v) \in T$.
\end{Lemma}
{\bf Proof:} $\Rightarrow$ By a slight abuse of notation consider $f$
to be a function on $\R \times {\cal U}$ satisfying $\partial
f/\partial t = 0$, where $t$ is the variable running along the $\R$
factor. Let $dt$ be the derivative of $t$ at $\left( f(\vec x_0), \vec
  x_0 \right)$, and let $df$ be the derivative of $f$ at $\left(
  f(\vec x_0), \vec x_0\right)$, then $T = \ker (dt - df).$

$\Leftarrow$ Let $(e_0, e_i)$ denote the standard basis of ${\Bbb R}
\times {\Bbb R}^n$ and let $(f^0, f^i)$ be the corresponding dual
basis.  Consider any $\alpha \in ({\Bbb R} \times {\Bbb R}^n)^*$ such
that $T = \ker \alpha$, thus $\alpha$ can be written in the form
$\alpha_r f^r$ (summation convention); note that $\alpha \neq 0$ since
codim $T = 1$.  Let $\vec v \in {\Bbb R}^n$ such that $\sum(v^i)^2 =
1$ and let $\epsilon_i$ be any sequence converging to zero; consider
the sequence $(f(\vec x _0 + \epsilon _i \vec v), \vec x_0 +
\epsilon_i \vec v) \rightarrow (f(\vec x_0), \vec x_0).$ Since $f$ is
Lipschitz continuous we have  $| f(\vec x_0 + \epsilon_i \vec v) -
f(\vec x_0)| \leq L \epsilon_i$ and compactness of $[-L, L]$
implies that there exists a subsequence $\epsilon_{i{_j}}$ such that
$$\left({f(\vec x_0 + \epsilon_{i{_j}} \vec v) - f(\vec x_0)\over
    \epsilon_{i{_j}}}, \vec v \right)$$ converges to $(v^0, \vec
v)$.  By hypothesis $(v^0, \vec v) \in T$, thus $\alpha_0 v^0 +
\alpha_i v^i = 0$. Note that $\alpha_0 v^0 = 0$ implies $\alpha_iv^i =
0$ and, hence, $\alpha_i = 0$ by arbitrariness of $v^i$.  It follows
that we can always normalize $\alpha$ so that $\alpha_0 = 1$ and we
get $v^0 = - \alpha_i v^i$.  We thus have
\begin{eqnarray}
  \lim_{j \rightarrow \infty} {f(\vec x_0 + \epsilon_{i{_j}} \vec v) -
    f(\vec x_0)\over \epsilon_{i{_j}}} &=& - \alpha_iv^i.  \label{1}
\end{eqnarray}
As the right--hand--side of \eq{1} does not depend upon the sequence
$\epsilon_i$, we must actually have
$$\lim_{\epsilon \rightarrow 0} {f(\vec x _0 + \epsilon \vec v) -
  f(\vec x_0) \over \epsilon} = - \alpha_iv^i.$$ This can be rewritten
as
$$f(\vec x_0 + \epsilon \vec v) = f(\vec x_0) -\epsilon \alpha_iv^i +
o(\epsilon), $$ which is what had to be established. \hfill$\Box$
\bpar Lemma \ref{L1} allows us to give a simple proof of the main
result of Beem and Kr\'olak \cite{BK2}; recall that ${\cal N}_p(\cal
H)$ denotes \cite{ChGalloway} the set of null semi-tangents at $p$ to
a Cauchy horizon $\cal H$, {\em i.e.}, the set of vectors tangent to
some generator of $\cal H$ through $p$, $p$ being possibly (but not
necessarily) an end point of such a generator, oriented to the past
for future Cauchy horizons, and to the future for past Cauchy
horizons. We also normalize those generators to length one with
respect to some fixed auxiliary Riemannian metric.
\begin{Theorem}[Beem and Kr\'olak \cite{BK2}]\label{T1} 
  Let $p \in {\cal H}$ be such that $\#{\cal N}_p({\cal H}) = 1$. Then
  ${\cal H}$ is differentiable at $p$.
\end{Theorem}

Theorem \ref{T1} follows immediately from the following, somewhat more
general statement:
\begin{Theorem}\label{T2} 
  Let ${\cal H}$ be a topological hypersurface satisfying the
  following:
  \begin{enumerate}
  \item ${\cal H}$ is {\em locally achronal}, {\em i.e.}, for any $p
    \in {\cal H}$ there exists a neighborhood $\cal O$ of $p$ such
    that ${\cal H}\cap \cal O$ is achronal in the space--time $({\cal
      O}, g|_{\cal O})$.
\item  Every point $p$ of ${\cal H}$ is either an interior point of a
  null geodesic $\Gamma\subset {\cal H}$, or a future endpoint
  thereof. Such $\Gamma$'s will be called generators of $\cal H$.
  \end{enumerate}
  Then ${\cal H}$ is differentiable at every point $p$ which belongs
  to only one generator of ${\cal H}$.
\end{Theorem}

{\bf Proof}: Let $p_i \in \cal H$ be any sequence such that $p_i
\rightarrow p$, and let $\gamma _i \in {\cal N}_{p_i}({\cal H})$; we
have $\gamma _i \rightarrow \gamma \in {\cal N}_p(\cal H)$ ({\em cf.,
  e.g.}, \cite[Lemma 3.1]{ChGalloway}, together with the argument of
the proof of Proposition 3.3 there).  In normal coordinates centered
at $p$ we can write
$$p_i = p + d_iv_i,\qquad 0 \leq d_i \rightarrow 0,$$ where the length
of the $v_i$'s has been normalized to $1$ using some auxiliary
Riemannian metric $M$. For $\epsilon, \epsilon_i\ge 0$ let
$p_i(\epsilon_i) \in \cal H$, respectively $p(\epsilon) \in \cal H$,
denote the point lying an affine distance $\epsilon_i$, respectively
$\epsilon$, on the null generator of $ \cal H$ with semi-tangent
$\gamma_i,$ respectively $\gamma$. From $\gamma_i\to\gamma$ we have
$\gamma_i-\gamma = o(1)$, and from the fact that in normal coordinates
null geodesics through $p+d_i v_i$ at affine distance $\epsilon _i$
differ from straight lines by terms which are $o(d_i+\epsilon_i)$ we obtain 
\begin{eqnarray*} 
  p_i (\epsilon_i) &=& p + d_i v_i + \epsilon_i\gamma_i + o(d_i + \epsilon_i),
  \\ p(\epsilon) &=& p + \epsilon \gamma. \end{eqnarray*} Let $\eta$
be the Minkowski metric, consider the 
quantity
\begin{eqnarray}
  A &=& \eta(p_i(\epsilon_i) - p, p_i(\epsilon_i) - p) \nonumber \\ 
  &=& d_i^2 \eta(v_i, v_i) + 2d_i \epsilon_i \eta(v_i, \gamma) +
  o((d_i + \epsilon_i)^2).  \label{2}
\end{eqnarray}
Suppose that $v_i \rightarrow v$, and suppose, first, that $\eta(v, v)
< 0$. Equation \eq{2} with $\epsilon_i = 0$ gives $A < 0$ for $i$ large enough.
It follows that the coordinate line through $p$ and $p_i(0) = p_i$ is
timelike which contradicts achronality of $\cal H$, hence
\begin{eqnarray}
  \eta(v, v) &\geq& 0. \label{3}
\end{eqnarray}
Suppose, next, that $\eta(v, v) > 0$ and $\eta(v, \gamma) < 0.$ In
that case Equation \eq{2} with $\displaystyle{\epsilon_i = {\eta(v, v)\over
    \mid \eta(v, \gamma)\mid} d_i}$ gives $A < 0$.  It follows that
the coordinate line through $p$ and $p_i(\epsilon_i)$ is timelike
which is again impossible, so that
\begin{eqnarray}
  \eta(v, \gamma) &\geq& 0. \label{4}
\end{eqnarray}
If $\eta(v, v) = 0$, Equation \eq{2} with $\epsilon_i = d_i$ leads similarly to
\eq{4}.

To show that the inequality \eq{4} has to be an equality, consider
the coordinate lines starting at $p+v_i$ and ending at $p+\epsilon_i\gamma$:
$$
[0,1]\ni s \to \Gamma_i(s)=p+(1-s)d_iv_i + s\epsilon_i\gamma\ .
$$
On $\Gamma_i(s)$ we have
\begin{eqnarray*}
  \label{table}
  g(v_i,v_i) &=&g(v,v) +o(1) = \eta(v,v)+o(1)\ ,\\
 g(\gamma,v_i)  &=&g(\gamma,v) +o(1) = \eta(\gamma,v)+o(1)\ ,\\ 
g(\gamma,\gamma)  &=&g(\gamma,\gamma) +o(1) = o(1)\ ,
\end{eqnarray*}
which implies
\be
g(\frac{d\Gamma_i}{ds},\frac{d\Gamma_i}{ds}) = d_i^2\eta(v,v) -
2\epsilon_id_i \eta(v,\gamma) + o((d_i+\epsilon_i)^2) \ .
\label{5}
\ee
Note that Equation \eq{5} differs from Equation \eq{2} only by the sign of the
$\eta(v_i, \gamma)$ terms, so that a similar analysis shows that
$\Gamma_i$ will be timelike for $i$ large enough unless
$$\eta(v, \gamma) = 0.$$ It follows that $v\in T\equiv \gamma^\perp$.
The local achronality of $\cal H$ implies that $\cal H$ is Lipschitz,
and differentiability of $\cal H$ at $p$ follows now from Lemma
\ref{L1}.  \hfill $\Box$

\providecommand{\bysame}{\leavevmode\hbox to3em{\hrulefill}\thinspace}

\end{document}